\documentstyle[12pt]{article}
\makeatletter
\def\section{\@startsection {section}{1}{\z@}{3.5ex plus 1ex minus
 .2ex}{2.3ex plus .2ex}{\Large\bf}}
\def\eqnarray{%
 \stepcounter{equation}%
 \let\@currentlabel=\theequation
 \global\@eqnswtrue
 \global\@eqcnt\z@
 \tabskip\@centering
 \let\\=\@eqncr
 $$\halign to \displaywidth\bgroup\@eqnsel\hskip\@centering
 $\displaystyle\tabskip\z@{##}$&\global\@eqcnt\@ne
 \hfil$\displaystyle{{}##{}}$\hfil
 &\global\@eqcnt\tw@$\displaystyle\tabskip\z@{##}$\hfil
 \tabskip\@centering&\llap{##}\tabskip\z@\cr}
\def\maketitle{\par
 \begingroup
 \def\thefootnote{\fnsymbol{footnote}}
 \def\@makefnmark{\hbox
 to 0pt{$^{\@thefnmark}$\hss}}
 \if@twocolumn
 \twocolumn[\@maketitle]
 \else \newpage
 \global \@topnum\z@ \@maketitle \fi
 \thispagestyle{empty}
 \@thanks
 \newpage
 \endgroup
 \setcounter{footnote}{0}
 \let\maketitle\relax
 \let\@maketitle\relax
 \gdef\@thanks{}\gdef\@author{}\gdef\@title{}
 \gdef\@prepri{}\gdef\@address{}
 \let\thanks\relax}
\def\@maketitle{\newpage
\begin{flushright}
{\em Ochanomizu University}
\hfill
\@prepri \\
\@date
\end{flushright}
 \vskip 2em
 \begin{center}
 {\LARGE \@title \par} \vskip 2em
 {\large \lineskip .5em
 \begin{tabular}[t]{c}\@author
 \end{tabular}\par}
 \vskip 1.5em
 {\em \begin{tabular}[t]{c}\@address
 \end{tabular}\par}
 \end{center}
 \vskip 1.5em
 \vfill
\@ifundefined{@abst}{}{
 \small
 \begin{center}
 {\bf Abstract\vspace{-.5em}\vspace{0pt}}
 \end{center}
 \begin{quotation} \@abst \end{quotation}
 \vfill
 \begin{center}
 Submitted to {\it Physics Letters} {\bf B}
 \end{center}
 \gdef\@abst{}}}
\long\def\abst#1{\long\gdef\@abst{#1}}
\def\prepri#1{\gdef\@prepri{OCHA-PP-#1}}
\def\address#1{\gdef\@address{#1}}
\makeatother

\begin{document}
\prepri{46\\YCCP-9408}
\date{November 5, 1994}
\title{Majorana neutrino versus Dirac neutrino \\
       in ${\rm e}^{+}{\rm e}^{-}\! \rightarrow {\rm W}^{+}{\rm W}^{-}$ \\
       through radiative corrections}
\author{Y.~Katsuki, M.~Marui, R.~Najima$^{*}$,\\
        J.~Saito$^{**}$ and A.~Sugamoto}
\address{Department of Physics, Faculty of Science,
         Ochanomizu University, \\
         Bunkyo-ku Tokyo 112, Japan \\
         \\
         $^{*}$
         Department of General Education,
         Yokohama College of Comerce, \\
         Kanagawa 203, Japan \\
         \\
         $^{**}$
         Suzugamine Women's College, \\
         Hiroshima 733, Japan}
\abst{Radiative corrections to
      ${\rm e}^{+}{\rm e}^{-}\! \rightarrow {\rm W}^{+}{\rm W}^{-}$
      from Majorana neutrinos are studied in the context
      of the see-saw mechanism.
      Focusing on the effects of the fourth generation neutrinos,
      we calculate W-pair form factors, the differential cross sections
      and the forward-backward asymmetries for the polarized electrons at
      one-loop level.
      The behaviour of the form factors at the threshold of Majorana
      particle pair productions
      is found to differ from that of Dirac particle pair productions.
      In the cross section for unpolarized electrons,
      the radiative corrections, depending on the mass parameters of
      the see-saw mechanism,
      are found to be $\sim 0.5\%$ at the energy range of the LEP200 and
      the next generation linear colliders.}
\maketitle
Precision experiments can probe new physics beyond the standard model,
because heavy particles do not necessarily decouple in low-energy processes
\cite{KL}. In the processes
${\rm e}^{+}{\rm e}^{-}\! \rightarrow {\rm f}\overline{{\rm f}}$
observed at the current colliders (LEP, SLC), the dominant effects of heavy
particles appear only in the corrections to the self erergies of gauge bosons.
These effects can be summarized into three parameters $S$, $T$ and $U$
\cite{PT} or $\epsilon _{1}$, $\epsilon _{2}$ and $\epsilon _{3}$ \cite{AB},
which have been estimated in various models. On the other hand , in the
process ${\rm e}^{+}{\rm e}^{-}\! \rightarrow {\rm W}^{+}{\rm W}^{-}$
which is the most important process in the LEP200 experiments, the effects
of heavy particles appear not only through the self energies of gauge bosons
but also through the trilinear gauge vertices
AW$^{+}$W$^{-}$ and ZW$^{+}$W$^{-}$.
Radiative corrections to this process in the standard model with heavy Dirac
fermions or heavy scalars have already been studied
\cite{APLS}. In this letter we investigate
radiative corrections to
${\rm e}^{+}{\rm e}^{-}\! \rightarrow {\rm W}^{+}{\rm W}^{-}$ from heavy
Majorana neutrinos in the context of the see-saw mechanism \cite{See-Saw}.
Focusing on the fourth generation neutrinos, we calculate W-pair form factors
and the differential cross sections for the polarized electrons and W bosons
at one-loop level.

We consider a model containing the fourth generation leptons having no
mixing with other generations \cite{HP}. Recent measurements at LEP and SLC
do not rule out the existence of the fourth neutrino but constrain its mass
to be greater than $45$ GeV. We assume the following mass term
for the neutrinos
\[
{\cal L}_{{\rm mass}}
  =-\frac{1}{2}\overline{(N_{{\rm L}}\; N_{{\rm R}}^{{\rm c}})}
       \left( \begin{array}{cc}
                0           & m_{{\rm D}} \\
                m_{{\rm D}} & M_{{\rm R}}
              \end{array} \right)
       \left( \begin{array}{c}
                N_{{\rm L}}^{{\rm c}} \\ N_{{\rm R}}
              \end{array} \right)
     + \mbox{h.c.},
\]
where $N^{{\rm c}}=C\overline{N}^{{\rm T}}$ is the charge-conjugated state of
$N$. We expect that the Dirac masses of neutrinos $m_{{\rm D}}$ generated by
the vacuum expectation value of ordinary Higgs bosons are comparable to
the masses
of their charged lepton partners $M_{\rm E}$. On the other hand, the
right-handed Majorana mass $M_{{\rm R}}$ originates from \lq \lq beyond the
standard model". In order to get positive mass eigenvalues, we diagonalize the
mass matrix and perform the chiral transformation. The result is given by
\[
{\cal L}_{{\rm mass}}
  =-\frac{1}{2}\overline{(N_{1}\; N_{2})}
      \left( \begin{array}{cc}
                M_{1} & 0    \\
                0     & M_{2}
             \end{array} \right)
      \left( \begin{array}{c} N_{1} \\ N_{2} \end{array} \right), \\
\]
where
\begin{eqnarray*}
 \left( \begin{array}{c} N_{1} \\ N_{2} \end{array} \right) & = &
  \left( \begin{array}{cc}
            i\gamma _{5}c & -i\gamma _{5}s \\
            s             & c
         \end{array} \right)
  \left( \begin{array}{c}
            N_{{\rm L}} + N_{{\rm L}}^{{\rm c}} \\
            N_{{\rm R}} + N_{{\rm R}}^{{\rm c}}
         \end{array} \right), \nonumber \\
 M_{1,2} & = &
  \frac{1}{2}(\sqrt{M_{{\rm R}}^{2}+4m_{{\rm D}}^{2}} \mp M_{{\rm R}}),
\\
 \tan \theta & = & \frac{M_{1}}{m_{{\rm D}}}
               = (\frac{M_{2}}{m_{{\rm D}}})^{-1} \qquad (M_{1}<M_{2}),
\end{eqnarray*}
where $s=\sin \theta, c=\cos \theta $.
In the fourth generation the smaller mass eigenvalue $M_{1}$ should be greater
than $45$ GeV. Moreover, the Dirac mass $m_{{\rm D}}$ is bounded above because
of the requirement of triviality; $m_{{\rm D}}$ is of the order of the weak
scale. Consequently, it turns out that the Majorana mass $M_{{\rm R}}$ cannot
be so large, and the two mass eigenvalues $M_{1}$ and $M_{2}$ are also the same
of order of magnitude, {\em i.e.}, the see-saw is almost balancing in the
heavy neutrino sector. In this case, the existence of the right-handed
Majorana mass $M_{{\rm R}}$ affects various weak processes through the large
left-right mixing. Contributions to $S$ and $T$ parameters in this model have
already been calculated and turned out to be negative when
$\tan \theta \sim 0.4$ \cite{BS,GT}. We will pay attention to this Majorana
mass region, because the precision mesurements at LEP favor negative values for
$S$ and $T$ \cite{Neg-ST}.

With the self-conjugacy of the majorana fields $N_{1}$ and $N_{2}$
under the charge conjugation operation,
the fermion currents can be written in terms of mass-eigenstates,
\begin{eqnarray}
 J_{\mu }^{Q} & = &
  Q^{\rm E}\overline{E}\gamma _{\mu }E,
\nonumber \\
 J_{\mu }^{3} & = &
  I_{3}^{\rm E}\overline{E}\gamma _{\mu }\frac{1-\gamma _{5}}{2}E
  +I_{3}^{\rm N}\left(
   -\frac{1}{2}c^{2}\overline{N_{1}}\gamma _{\mu }\gamma _{5}N_{1}
   -\frac{1}{2}s^{2}\overline{N_{2}}\gamma _{\mu }\gamma _{5}N_{2}
   +isc\overline{N_{2}}\gamma _{\mu }N_{1}\right),
\nonumber \\
 J_{\mu }^{+} & = &
   \frac{1}{\sqrt{2}}
    \left(-ic\overline{N_{1}}+s\overline{N_{2}}\right)
     \gamma _{\mu }\frac{1-\gamma _{5}}{2}E,
\nonumber \\
 J_{\mu }^{-} & = &
   \frac{1}{\sqrt{2}}
    \overline{E}\gamma _{\mu }\frac{1-\gamma _{5}}{2}
    \left(icN_{1}+sN_{2}\right),
\label{eq:currents}
\end{eqnarray}
where $E$ refers to the
charged lepton field. We note that the Z coupling to
neutrinos with the same masses is axial vector, while it becomes a vector
coupling when mixing the neutrinos with different masses.

We now consider the radiative corrections to
${\rm e}^{+}{\rm e}^{-}\! \rightarrow {\rm W}^{+}{\rm W}^{-}$
at one-loop level.
Because the mixing between heavy leptons and an electron is probably small
and negligible, if it exist,
the radiative corrections appear dominantly in the self energies and
the corrections to the
trilinear vertex corrections of gauge bosons. These corrections belong to
\lq \lq oblique" ones \cite{KL}.
A renormalization
program for oblique corrections is proposed by Kennedy and Lynn on the
concept of effective lagrangian \cite{KL}. The extension of their formalism
to the process of
${\rm e}^{+}{\rm e}^{-}\! \rightarrow {\rm W}^{+}{\rm W}^{-}$ is given by
Ahn, Peskin, Lynn and Selipsky \cite{APLS}.
They express the s-channel amplitude at one-loop level for on-shell W bosons
with running coupling constant $e_{*}$, the running wave-function
renormalization
constant of W boson $Z_{{\rm W}*}$ \cite{KL} and the vertex function
$\Gamma ^{\mu \alpha \beta }$ into which A and Z vertices are combined:
\[
{\cal M}_{s}^{{\rm 1-loop}}=
     -\frac{ie_{*}^{2}}{P^{2}}Z_{{\rm W}*}\overline{v}\gamma _{\mu }u
     \Gamma ^{\mu \alpha \beta }
     \varepsilon _{\alpha }^{*}(q)\varepsilon _{\beta }^{*}(\overline{q}),
\]
where $P,\, q$ and $\overline{q}$ are momenta of V, ${\rm W}^{-}$ and
${\rm W}^{+}$,
$u$ and $v$ are spinors of ${\rm e}^{-}$ and ${\rm e}^{+}$ and
$\varepsilon _{\alpha }(q)$ and $\varepsilon _{\beta }(\overline{q})$
refer to the polarization-vectors of ${\rm W}^{-}$ and ${\rm W}^{+}$,
respectively.

Here, we define the modified vertex function
$\overline{\Gamma }^{\mu \alpha \beta }
 =Z_{{\rm W}*}\Gamma ^{\mu \alpha \beta }$
and express it in terms of W-pair form factors $f_{i}^{\rm V=A,Z}$
and canonical Lorentz structures $T_{i}$, following
Hagiwara {\em et al.} \cite{HPZH}.
(In ref.~\cite{APLS}, $Z_{{\rm W}*}$ is omitted in the definition of
W-pair form factors and is multiplied to the cross sections afterward
as the overall factor.)
\[
\overline{\Gamma }^{\mu \alpha \beta }
 =\sum_{i=1}^{7}(Qf_{i}^{{\rm A}}
  +\frac{I_{3}-s_{*}^{2}Q}{s_{*}^{2}}\frac{P^{2}}{P^{2}-m_{{\rm Z}}^{2}}
  f_{i}^{{\rm Z}})T_{i}^{\mu \alpha \beta },
\]
\begin{eqnarray*}
T_{1}^{\mu \alpha \beta } & = & (q-\overline{q})^{\mu }g^{\alpha \beta },
\qquad \qquad
T_{2}^{\mu \alpha \beta }=-\frac{1}{m_{{\rm W}}^{2}}
                          (q-\overline{q})^{\mu }P^{\alpha }P^{\beta },
\\
T_{3}^{\mu \alpha \beta } & = & P^{\alpha }g^{\mu \beta }
                                -P^{\beta }g^{\mu \alpha },
\qquad
T_{4}^{\mu \alpha \beta }=i(P^{\alpha }g^{\mu \beta }
                            +P^{\beta }g^{\mu \alpha }),
\\
T_{5}^{\mu \alpha \beta } & = & i\epsilon ^{\mu \alpha \beta \rho }
                                 (q-\overline{q})_{\rho },
\qquad \quad
T_{6}^{\mu \alpha \beta }=-\epsilon ^{\mu \alpha \beta \rho }P_{\rho },
\\
T_{7}^{\mu \alpha \beta } & = &
  -\frac{1}{m_{{\rm W}}}(q-\overline{q})^{\mu }
  \epsilon ^{\alpha \beta \rho \sigma }P_{\rho }(q-\overline{q})_{\sigma },
\end{eqnarray*}
where $s_{*}$ is the running weak mixing angle defined by Kennedy and Lynn,
$Q$ and $I_{3}$ are the charge and the isospin of ${\rm e}^{-}$, respectively,
and we use the Minkowskian convention with $\epsilon _{0123}=1$.
Form factors can be written in terms of the self energies $\Pi _{XY}$ and
the vertex corrections $\Sigma _{X}^{i}$
\begin{eqnarray}
f_{1}^{{\rm A}} & = &
 1+g_{*}^{2}\bigl[\Sigma _{Q}^{1}+\Pi' _{11}\bigr],
\nonumber \\
f_{1}^{{\rm Z}} & = &
 1+\frac{g_{*}^{2}}{c_{*}^{2}}
  \bigl[(\Sigma _{3}^{1}-s_{*}^{2}\Sigma _{Q}^{1})+c_{*}^{2}\Pi' _{11}\bigr]
  +\frac{M_{{\rm Z}*}^{2}-m_{{\rm Z}}^{2}}{P^{2}-m_{{\rm z}}^{2}},
\nonumber \\
f_{3}^{{\rm A}} & = &
 2+g_{*}^{2}\bigl[\Sigma _{Q}^{3}+2\Pi' _{11}\bigr],
\nonumber \\
f_{3}^{{\rm Z}} & = &
 2+\frac{g_{*}^{2}}{c_{*}^{2}}
  \bigl[(\Sigma _{3}^{3}-s_{*}^{2}\Sigma _{Q}^{3})+2c_{*}^{2}\Pi' _{11}\bigr]
  +2\frac{M_{{\rm Z}*}^{2}-m_{{\rm Z}}^{2}}{P^{2}-m_{{\rm z}}^{2}},
\nonumber \\
f_{i}^{{\rm A}} & = &
 g_{*}^{2}\bigl[\Sigma _{Q}^{i}\bigr],
\nonumber \\
f_{i}^{{\rm Z}} & = &
 \frac{g_{*}^{2}}{c_{*}^{2}}
  \bigl[(\Sigma _{3}^{i}-s_{*}^{2}\Sigma _{Q}^{i})\bigr],
\qquad \qquad i=2,\, 4,\, 6,\, 7,
\label{eq:form-factors}
\end{eqnarray}
where we adopt the notation of ref.~\cite{APLS},
$g_{*}$ and $M_{{\rm Z}*}$ is the running coupling constant and the running
Z boson's mass defined by Kennedy and Lynn, respectively, and
$\Pi '_{11}\equiv \Pi '_{11}(m_{\rm W}^{2})$.
These form factors, especially in the limit
of low energies, are controled by gauge or global symmetries
\cite{APLS,HPZH,H,AW}.

Using the currents in eq.(\ref{eq:currents}), we can now calculate
the one-loop corrections in eq.(\ref{eq:form-factors}) due to the leptons
in the fourth generation. First, we present the expressions for the self
energies
\begin{eqnarray}
16\pi ^{2}\Pi _{QQ} & = &
 -(Q^{\rm E})^{2}8P^{2}\bigl[\frac{1}{6}\overline{\epsilon }^{-1}
 - {\bf b_{3}}(P^{2},M_{\rm E}^{2},M_{\rm E}^{2})\bigr],
\nonumber \\
16\pi ^{2}\Pi _{3Q} & = &
 -(Q^{\rm E}I_{3}^{\rm E})4P^{2}\bigl[\frac{1}{6}\overline{\epsilon }^{-1}
 - {\bf b_{3}}(P^{2},M_{\rm E}^{2},M_{\rm E}^{2})\bigr],
\nonumber \\
16\pi ^{2}\Pi _{33} & = &
 -(I_{3}^{\rm E})^{2}
 \bigl[4P^{2}\{\frac{1}{6}\overline{\epsilon }^{-1}
               -{\bf b_{3}}(P^{2},M_{\rm E}^{2},M_{\rm E}^{2})\}
      -2M_{\rm E}^{2}\{\overline{\epsilon }^{-1}
               +{\bf b_{0}}(P^{2},M_{\rm E}^{2},M_{\rm E}^{2})\}\bigr]
\nonumber \\
 & & \mbox{}
 -(I_{3}^{\rm N}c^{2})^{2}
 \bigl[4P^{2}\{\frac{1}{6}\overline{\epsilon }^{-1}
               -{\bf b_{3}}(P^{2},M_{1}^{2},M_{1}^{2})\}
      -4M_{1}^{2}\{\overline{\epsilon }^{-1}
               +{\bf b_{0}}(P^{2},M_{1}^{2},M_{1}^{2})\}\bigr]
\nonumber \\
 & & \mbox{}
 -(I_{3}^{\rm N}s^{2})^{2}
 \bigl[4P^{2}\{\frac{1}{6}\overline{\epsilon }^{-1}
               -{\bf b_{3}}(P^{2},M_{2}^{2},M_{2}^{2})\}
      -4M_{2}^{2}\{\overline{\epsilon }^{-1}
               +{\bf b_{0}}(P^{2},M_{2}^{2},M_{2}^{2})\}\bigr]
\nonumber \\
 & & \mbox{}
 -2(I_{3}^{\rm N}c^{2})(I_{3}^{\rm N}s^{2})
 \bigl[4P^{2}\{\frac{1}{6}\overline{\epsilon }^{-1}
               -{\bf b_{3}}(P^{2},M_{1}^{2},M_{2}^{2})\}
\nonumber \\
 & & \qquad \qquad \qquad \qquad
      -(M_{1}-M_{2})^{2}\{\overline{\epsilon }^{-1}
               +{\bf b_{0}}(P^{2},M_{1}^{2},M_{2}^{2})\}
\nonumber \\
 & & \qquad \qquad \qquad \qquad
       +(M_{1}^{2}-M_{2}^{2})
        \{{\bf b_{1}}(P^{2},M_{1}^{2},M_{2}^{2})
          -{\bf b_{1}}(P^{2},M_{2}^{2},M_{1}^{2})\}\bigr],
\nonumber \\
16\pi ^{2}\Pi _{11} & = &
 -c^{2}\bigl[2P^{2}\{\frac{1}{6}\overline{\epsilon }^{-1}
               -{\bf b_{3}}(P^{2},M_{1}^{2},M_{\rm E}^{2})\}
              -\frac{1}{2}(M_{\rm E}^{2}+M_{1}^{2})\overline{\epsilon }^{-1}
\nonumber \\
 & & \qquad \qquad \quad
         +M_{1}^{2}{\bf b_{1}}(P^{2},M_{1}^{2},M_{\rm E}^{2})
         +M_{\rm E}^{2}{\bf b_{1}}(P^{2},M_{\rm E}^{2},M_{1}^{2})\}\bigr]
\nonumber \\
 & & \mbox{}
 -s^{2}\bigl[2P^{2}\{\frac{1}{6}\overline{\epsilon }^{-1}
               -{\bf b_{3}}(P^{2},M_{2}^{2},M_{\rm E}^{2})\}
              -\frac{1}{2}(M_{\rm E}^{2}+M_{2}^{2})\overline{\epsilon }^{-1}
\nonumber \\
 & & \qquad \qquad \quad
         +M_{2}^{2}{\bf b_{1}}(P^{2},M_{2}^{2},M_{\rm E}^{2})
         +M_{\rm E}^{2}{\bf b_{1}}(P^{2},M_{\rm E}^{2},M_{2}^{2})\}\bigr],
\label{eq:pi}
\end{eqnarray}
where $\overline{\epsilon }^{-1}$ is the divergence of the dimentional
regulatization,
$\overline{\epsilon }^{-1}=2/(4-D)-\gamma+\ln 2\pi$.
The vertex corrections can be written as follows
\begin{eqnarray}
\Sigma_{Q} & = &
 \frac{1}{2}Q^{\rm E}c^{2}\bigl[
  H(P^{2},M_{\rm E}^{2},M_{\rm E}^{2},M_{1}^{2})
  -G(P^{2},M_{\rm E}^{2},M_{\rm E}^{2},M_{1}^{2})\bigr]
\nonumber \\
 & & \mbox{}
 +\frac{1}{2}Q^{\rm E}s^{2}\bigl[
  H(P^{2},M_{\rm E}^{2},M_{\rm E}^{2},M_{2}^{2})
  -G(P^{2},M_{\rm E}^{2},M_{\rm E}^{2},M_{2}^{2})\bigr],
\nonumber \\
\Sigma_{3} & = &
 \frac{1}{2}I_{3}^{\rm E}\bigl[
  c^{2}H(P^{2},M_{\rm E}^{2},M_{\rm E}^{2},M_{1}^{2})
  +s^{2}H(P^{2},M_{\rm E}^{2},M_{\rm E}^{2},M_{2}^{2})\bigr]
\nonumber \\
 & & \mbox{}
 -\frac{1}{2}I_{3}^{N}\bigl[
  c^{4}\{\tilde{H}(P^{2},M_{1}^{2},M_{1}^{2},M_{\rm E}^{2})
        +\tilde{G}(P^{2},M_{1}^{2},M_{1}^{2},M_{\rm E}^{2})\}
\nonumber \\
 & & \qquad \qquad
 +s^{4}\{\tilde{H}(P^{2},M_{2}^{2},M_{2}^{2},M_{\rm E}^{2})
        +\tilde{G}(P^{2},M_{2}^{2},M_{2}^{2},M_{\rm E}^{2})\}
\nonumber \\
 & & \qquad \qquad
 +2s^{2}c^{2}\{\tilde{H}(P^{2},M_{1}^{2},M_{2}^{2},M_{\rm E}^{2})
              -\tilde{G}(P^{2},M_{1}^{2},M_{2}^{2},M_{\rm E}^{2})\}\bigr].
\label{eq:sigma}
\end{eqnarray}
The functions $H$ and $G$ in the above expression are
\begin{eqnarray}
16\pi ^{2}H(P^{2},m_{1}^{2},m_{2}^{2},m_{3}^{2}) & = &
 T_{0}\bigl[-\frac{2}{3}\overline{\epsilon }^{-1}
  +{\bf c_{0}}+{\bf c_{1}}
  -\frac{m_{\rm W}^{2}}{\mu ^{2}}({\bf c_{4}}-{\bf c_{5}})
  -\frac{P^{2}}{\mu ^{2}}({\bf c_{6}}+{\bf c_{7}})\bigr]
\nonumber \\
 & & \mbox{}
 +T_{2}\bigl[-4\frac{m_{\rm W}^{2}}{\mu ^{2}}{\bf c_{7}}\bigr]
\nonumber \\
 & & \mbox{}
 +T_{3}\bigl[{\bf c_{0}}-3{\bf c_{1}}
  -\frac{m_{\rm W}^{2}}{\mu ^{2}}(2{\bf c_{3}}-5{\bf c_{4}}+3{\bf c_{5}})
  +\frac{P^{2}}{\mu ^{2}}({\bf c_{6}}+3{\bf c_{7}})\bigr]
\nonumber \\
 & & \mbox{}
 +T_{5}\bigl[{\bf c_{0}}-3{\bf c_{1}}
  +\frac{m_{\rm W}^{2}}{\mu ^{2}}({\bf c_{4}}-{\bf c_{5}})
  -\frac{P^{2}}{\mu ^{2}}({\bf c_{6}}-{\bf c_{7}})\bigr],
\nonumber \\
16\pi ^{2}G(P^{2},m_{1}^{2},m_{2}^{2},m_{3}^{2}) & = &
 (T_{0}-T_{3}+T_{5})
   \bigl[\frac{m_{1}m_{2}}{\mu ^{2}}({\bf c_{2}}-{\bf c_{3}})\bigr].
\label{eq:HG}
\end{eqnarray}
Here $T_{0}\equiv T_{1}+2T_{3}$ is the only Lorentz structure at tree
level, $\mu $ is a reference point and the ${\bf c_{i}}$ have arguments
$(P^{2},m_{1}^{2},m_{2}^{2},m_{3}^{3})$.
We ommit gauge anomaly terms in eq.(\ref{eq:HG}), because they should cancel
themselves if we consider the full fourth generation.
The functions $\tilde{H}$ and $\tilde{G}$ are obtained by reversing the sign
of $T_{5}$ in eq.(\ref{eq:HG}). The reduced Passarino-Veltman functions
${\bf b_{i}}(P^{2},m_{1}^{2},m_{2}^{2})$ and
${\bf c_{i}}(P^{2},m_{1}^{2},m_{2}^{2},m_{3}^{3})$ in eq.(\ref{eq:pi}) and
(\ref{eq:HG}) \cite{APLS,PV} are defined as folows
\begin{eqnarray*}
 [{\bf b_{0}},{\bf b_{1}},{\bf b_{3}}] & = &
  \int \! dxdy \delta (1-x-y)\ln (D_{2}/\mu ^{2})[-1,x,xy],
\\ \mbox{}
 [{\bf c_{0}},{\bf c_{1}}] & = &
  \int \! dxdydz \delta (1-x-y-z)\ln (D_{3}/\mu ^{2})[1,z],
\\ \mbox{}
 [{\bf c_{2}},{\bf c_{3}},{\bf c_{4}},{\bf c_{5}},{\bf c_{6}},{\bf c_{7}}]&=&
 \int \! dxdydz \delta (1-x-y-z) (\mu ^{2}/D_{3})
\\
 & & \quad
 \times [1,z,z^{2},z^{3},xy,xyz],
\end{eqnarray*}
\begin{eqnarray}
 D_{2}&=&-xyP^{2}+xm_{1}^{2}+ym_{2}^{2}-i\epsilon ,
\nonumber  \\
 D_{3}&=&-z(1-z)m_{\rm W}^{2}-xyP^{2}
  +xm_{1}^{2}+ym_{2}^{2}+zm_{3}^{2}-i\epsilon .
\label{eq:bc}
\end{eqnarray}

We note that Lorentz structures $T_{4}$, $T_{6}$ and $T_{7}$ disappear
in eq.(\ref{eq:HG}). The W-pair form factors corresponding to these Lorentz
structures are $CP$ variant \cite{HPZH}, and hence should vanish explicitly
in $CP$ conserving models such as we consider here.
We also note that the
threshold behaviour of the form factors for ${\rm N}_{1}$s'
(${\rm N}_{2}$'s) pair production
and for ${\rm N}_{1}$-${\rm N}_{2}$ production are different.
In the self energies, this difference can be seen easily from eq.(\ref{eq:pi}),
as a differnce in the power of $\beta _{ij}$;
$\beta _{ij}\equiv [1-(M_{i}+M_{j})^{2}/P^{2}]^{1/2}$.
The leading terms of $\Pi _{33}$ at ${\rm N}_{1}$s' (${\rm N}_{2}$'s)
pair production threshold
and
at ${\rm N}_{1}$-${\rm N}_{2}$ production threshold
are proportional to $\beta _{11(22)}^{3}$ and
$\beta _{12}$, respectively \cite{M}.
We will discuss the threshold behaviour of the form factors containing
the vertex corrections afterward.

Now, we peform the numerical calculations using the method in ref \cite{FSKO}.
We fix the following parameters to the
values at energy scale of $m_{{\rm Z}}$:
\[
\alpha ^{-1}_{*} = \frac{4\pi }{e_{*}^{2}} = 128.0, \qquad s_{*}^{2} = 0.2319,
\qquad
m_{{\rm Z}} = 91.187\; \mbox{[GeV]},
\]
where we neglect the running effects of $e_{*},\, s_{*}$ \cite{KL,PT}.
We also set $m_{{\rm W}} = 80.22$ [GeV].

The real parts of the corrections to W-pair form factors
$\Delta f_{i=1,2,3,5}^{\rm Z}$
from heavy neutrinos and charged
leptons are shown in fig.1(a) and 1(b),
where neutrinos are Dirac and Majorana particles, respectively.
The form factors $\Delta f_{2}^{\rm A,Z}$ are quite small,
as can be seen in fig.1, since these form factors correspond to the
dimension $6$ operators \cite{APLS,HPZH}.
Therefore the radiative corrections due to the heavy fermions mainly
appear in the other form factors.
We can see cusps in the $\Delta f_{1,3,5}^{\rm Z}$-lines at the
Dirac neutrino threshold in fig.1(a) and in the $\Delta f_{5}^{\rm Z}$-line at
${\rm N}_{1}$-${\rm N}_{2}$ threshold in fig.1(b). The enhancement at
${\rm N}_{1}$-${\rm N}_{2}$ threshold is very small because of the small
couplings including the mixing angle in eq.(1). On the other hand,
$\Delta f_{1,3,5}^{\rm Z}$-lines are not cusped at ${\rm N}_{1}$-${\rm N}_{1}$
threshold as seen in fig.1(b). In self energies, as mentioned above,
the difference of threshold behaviours depending on the neutrino types
has been checked analytically.
Numerical calculations including vertex corrections also suggest that
threshold behaviours are scalar-like for ${\rm N}_{1}$'s (${\rm N}_{2}$'s)
pair production, but Dirac fermion-like for ${\rm N}_{1}$-${\rm N}_{2}$
production. This is caused by the difference in the coupling types between
$Z\overline{N_{1}}N_{1}$ ($Z\overline{N_{2}}N_{2}$) and
$Z\overline{N_{1}}N_{2}$ in eq.(\ref{eq:currents}).
It is also helpful to recognize that the two Majorana particles of
the same kind (${\rm N}_{1}$-${\rm N}_{1}$ or ${\rm N}_{2}$-${\rm N}_{2}$) are
produced by the p-wave from the spin 1 photon or Z boson, in the same manner as
the scalars' pair production, whereas the different kinds of neutrinos
(${\rm N}_{1}$-${\rm N}_{2}$) are produced by the s-wave,
which happens in the ordinary Dirac fermions' pair production.
The difference appeared in the threshold behaviours may be understood as
$\beta ^{2l+1}$ for the $l$-th wave with the relative velocity $\beta $.

Next, we present the results on the differential cross sections
$d\sigma /d\cos \Theta $ and
the forward-backward asymmetries $A_{FB}$.
Since the vertex corrections, on which we concentrate in this paper, appear
only on s-channel gauge boson exchange diagrams, we plot $d\sigma /d\cos \Theta
$ and $A_{FB}$ for right-handed polarized electrons in fig.2 $\sim $ fig.4.
Hereafter we consider the full fourth generation with
degenerate quarks, and set $M_{\rm U}=M_{\rm D}=M_{\rm E}=500[\rm GeV]$, where
$M_{\rm U}$, $M_{\rm D}$ and $M_{\rm E}$ are masses of up-type quark,
down-type quark and charged lepton, respectively.
In fig.2, we plot the differential cross section $d\sigma /d\cos \Theta $
at scattering angle $\Theta = \frac{\pi }{2}$
in units of the point cross section $1R=4\pi \alpha ^{2}_{*}/3s$ with $s=P^{2}$
and $\alpha_{*} =1/128.0$.
We find that the differntial cross sections depend on the neutrino types, i.e.,
Majorana neutrino or Dirac neutrino, and their masses, although no
sizable peaks
can be seen except for the charged particles' pair production threshold.
The anguler distribution of the differntial cross section is shown in fig.3.
At any scattering angle, the heavy fermions make the correction less than
about $1\%$ at $\sqrt{s}=170$ [GeV] and
less than about $4\%$ at $\sqrt{s}=300$ [GeV].
In fig.4 we plot the forward-backward asymmetry $A_{FB}$ for
${\rm e}^{+}_{\rm L}{\rm e}^{-}_{\rm R}\! \rightarrow {\rm W}^{+}{\rm W}^{-}$
with unpolarized final W's,
in the angular range $|\cos \Theta |\le 0.98$. For this initial polarization
, $A_{FB}$
is controled by two form factors $f_{5}^{\rm A}$ and $f_{5}^{\rm Z}$ which
vanish at tree level. At one-loop level, these form factors also vanish when
we consider a full generation with degenerate Dirac fermions.
(Note that the functions $\tilde{H}$ and $\tilde{G}$ in eq.(\ref{eq:sigma})
are obtained by reversing the sign of $T_{5}$ in eq.(\ref{eq:HG}).)
Hence the non-vanishing $A_{FB}$ is solely due to the Majorana mass
$M_{\rm R}$ in line(B) or due to the Dirac mass spliting in line(D).

To discuss the observability of the radiative corrections at LEP200 or at the
next generation linear colliders, we estimate the cross section for
unpolarized electrons.
At tree level, the cross section $\sigma $ in the angular range
$|\cos \Theta |\le 0.98$ is $16.80$ [pb] at $\sqrt{s}=175$ [GeV] and
$11.45$ [pb] at $\sqrt{s}=300$ [GeV]. The effect of radiative corrections is
$0.6 \sim 0.7\%$ at both energies. At LEP200 with an integrated luminosity of
$500$ [pb$^{-1}$], it is difficult to measure the deviation from the tree level
value.
However, at the next generation linear colliders with
$\sqrt{s}=300\, [{\rm GeV}] \sim 1.5\, [{\rm TeV}]$ and an integrated
luminosity of $5\times 10^{4}$ [pb$^{-1}$], the deviation depend on the
neutrino types and their masses will be mesureable.

In summary, we have investigated the one-loop corrections due to the heavy
Majorana neutrinos in the process
${\rm e}^{+}{\rm e}^{-}\! \rightarrow{\rm W}^{+}{\rm W}^{-}$ which will be
seen in LEP200 and the next generation linear colliders.
We have estimated W-pair form factors in terms of reduced Passarino-Veltman
functions.
We have found that the threshold behaviour of these form factors is
scalar-like for a pair of
light neutrinos (${\rm N}_{1}$-${\rm N}_{1}$) or a pair of heavy neutrinos
(${\rm N}_{2}$-${\rm N}_{2}$) production, while that is Dirac fermion-like for
light and heavy neutrinos (${\rm N}_{1}$-${\rm N}_{2}$) production.
We also have estimated the differential cross section and the
forward-backward asymmetry, numerically,
and have found them depending on the Dirac mass and the mixing angle in the
see-saw mechanism even at low energies.
The deviation of the cross section from the tree level one will be measurable
at the next linear colliders.

\section*{Acknowledgments}
The authors would like to thank N. Nakazawa and K. Kato who offered
the computer program useful for performing Feynman Parameter integrals.
They are also grateful to T. Inami, C.S. Lim and I.Watanabe for discussions.
This work is in part supported by Grant-in-Aid for Scientific Research
from the Ministry of Education, Science and Culture (No. 05640336).

\newpage
\section*{Figure Captions}
\begin{description}
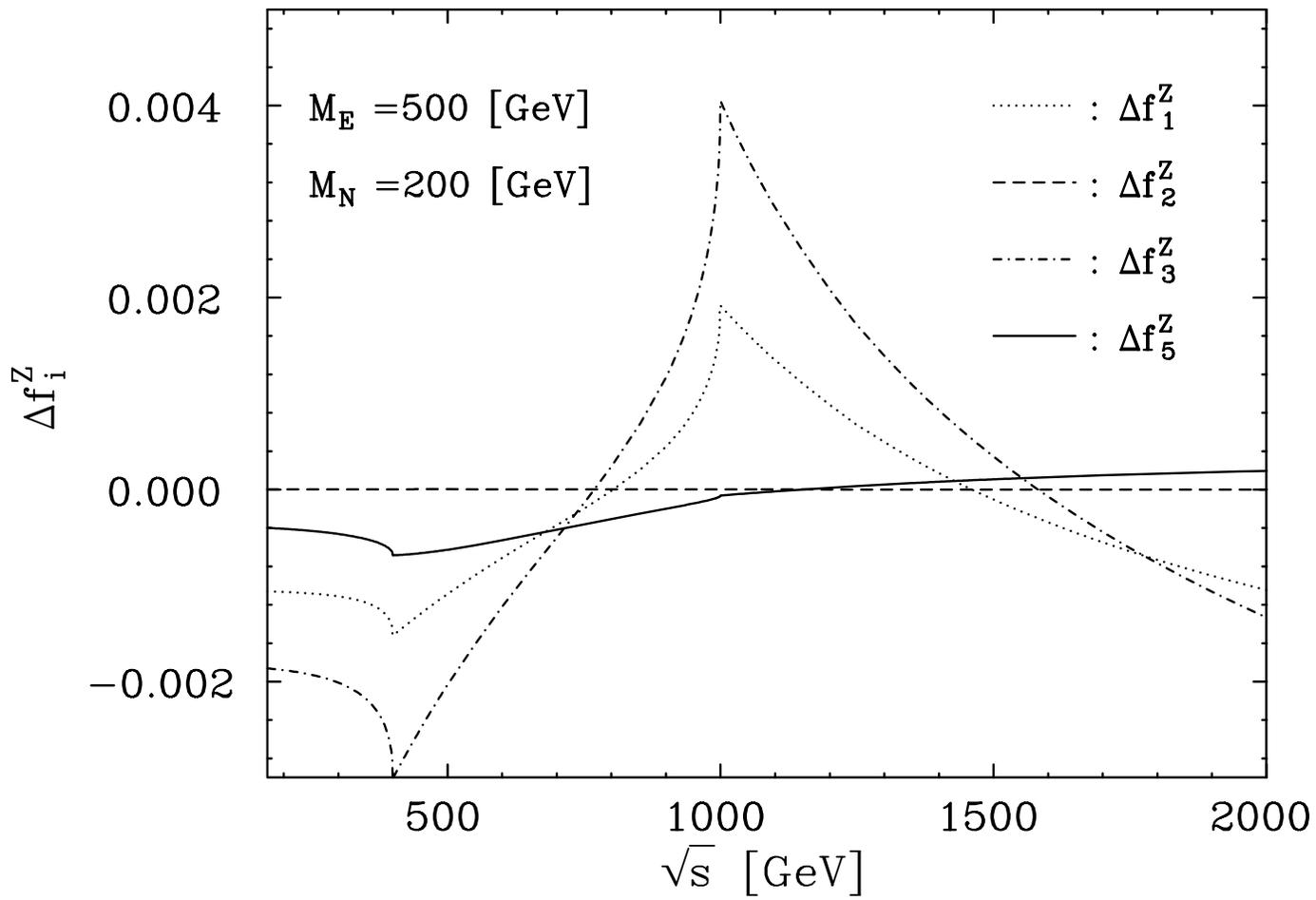
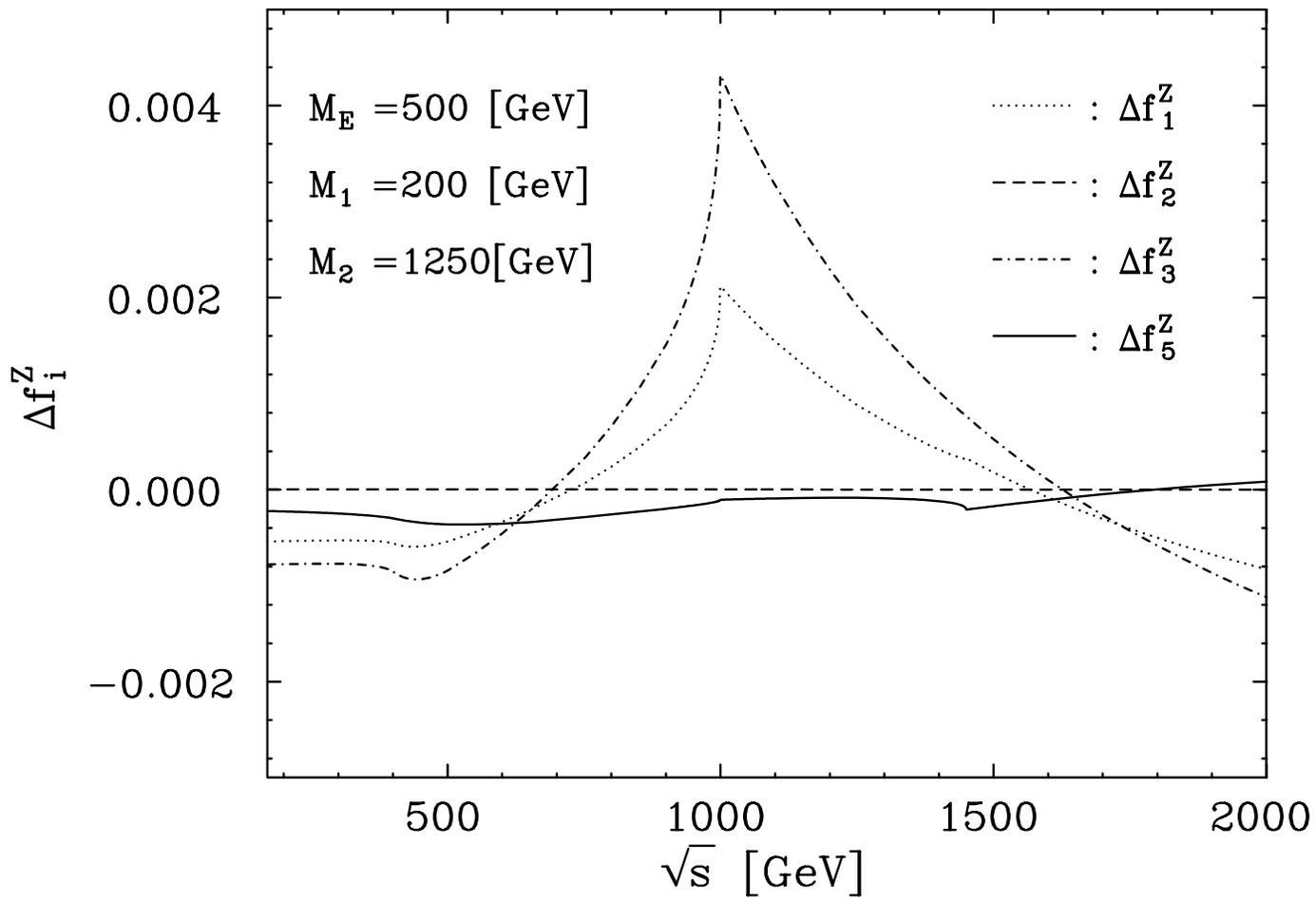
\item[fig.1]
 Correction $\Delta f^{\rm Z}_{i=1,2,3,5}$
 to W-pair form factors due to the fourth generation leptons
 as a function of $\sqrt{s}$. \\
 Fig.1(a) is given for the Dirac neutrino $\rm N$ with $M_{\rm N}=200$ [GeV].
 Fig.1(b) is given for the Majorana neutrinos ${\rm N}_{1}$ and ${\rm N}_{2}$
 with $M_{1}=200$ [GeV] and $M_{2}=1250$ [GeV]
 ($m_{\rm D}=500$ [GeV] and $\tan \theta =0.4$).
 We set the charged lepton mass $M_{\rm E}$ at $500$ [GeV] in the both figure.
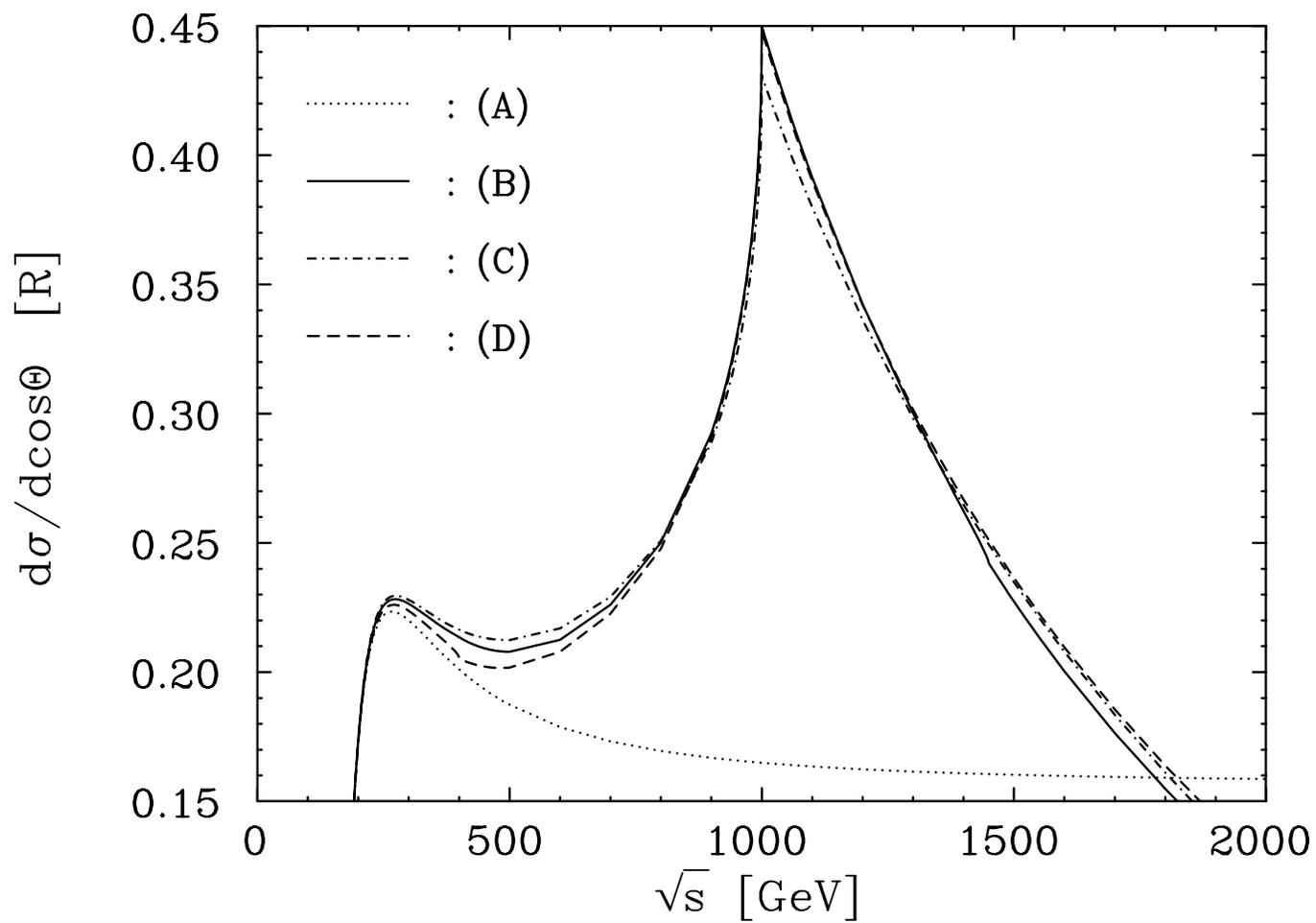
\item[fig.2]
 Differntial cross section $\frac{d\sigma }{d\cos \Theta }$ for
 ${\rm e}^{+}_{\rm L}{\rm e}^{-}_{\rm R}\! \rightarrow {\rm W}^{+}{\rm W}^{-}$
 at scattering angle $\Theta = \frac{\pi }{2}$ as a function of $\sqrt{s}$. \\
 The dotted line (A) shows the differntial cross section at tree level.
 The solid line (B), the dashed-and-dotted line (C) and the dashed line (D)
 include the corrections due to the quarks and the leptons of the fourth
 generation at one-loop level.
 The line (B) is depicted
 for the Majorana neutrinos ${\rm N}_{1}$ and ${\rm N}_{2}$
 with $M_{1}=200$ [GeV] and $M_{2}=1250$ [GeV].
 The line (C) and the line (D) are shown for the Dirac neutrino $\rm N$ with
 $M_{\rm N} = 500$ [GeV] and $200$ [GeV], respectively.
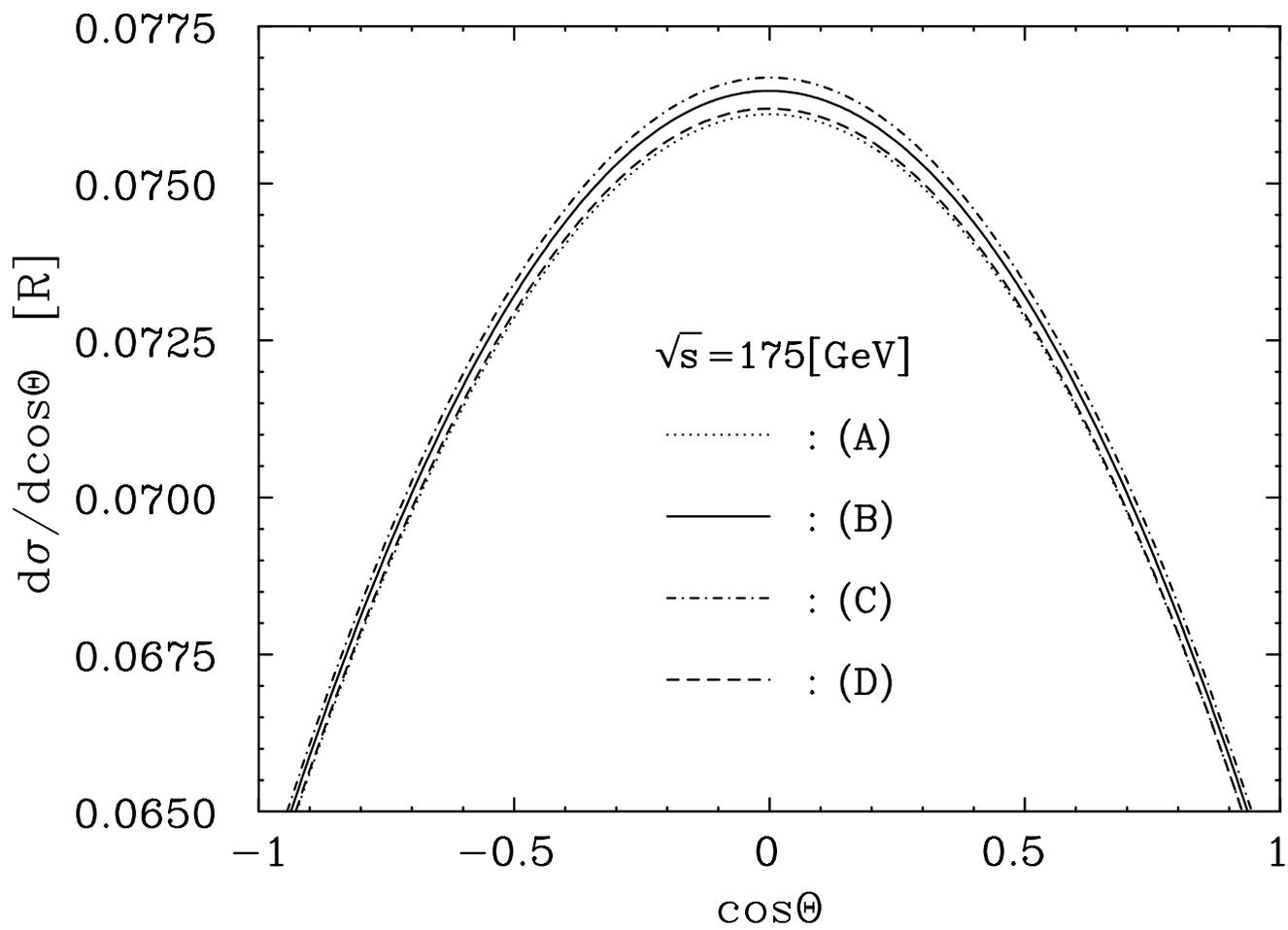
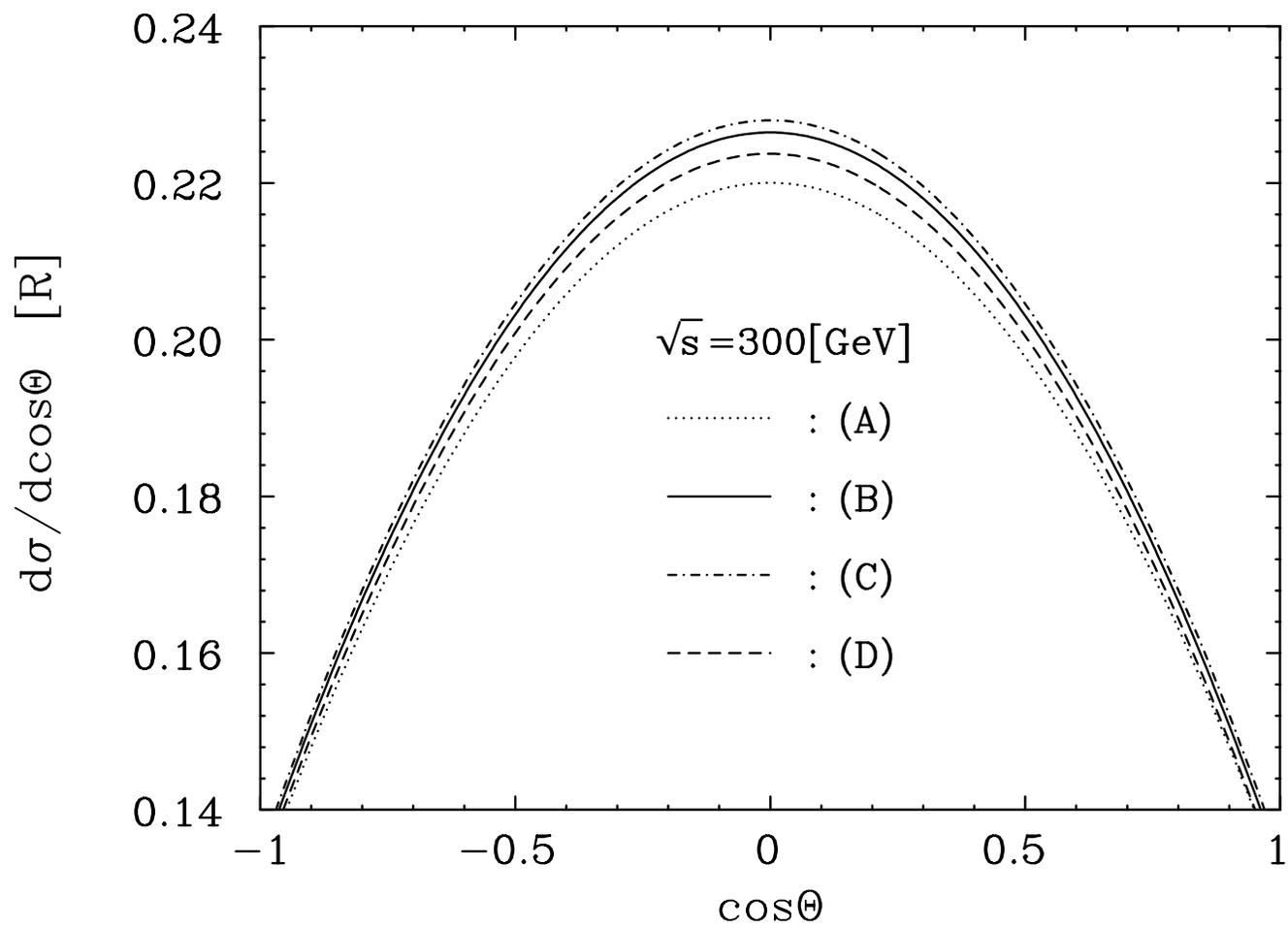
\item[fig.3]
 Angular distribution of the differntial cross section
 $d\sigma /d\cos \Theta $ for
 ${\rm e}^{+}_{\rm L}{\rm e}^{-}_{\rm R}\! \rightarrow {\rm W}^{+}{\rm W}^{-}$.
 Fig.3(a) and 3(b) are at $\sqrt{s}=175$ [GeV] and 300 [GeV], respectively.
 The values of the masses in (A),(B),(C) and (D) are the same as those in
 the corresponding curves in fig.2.
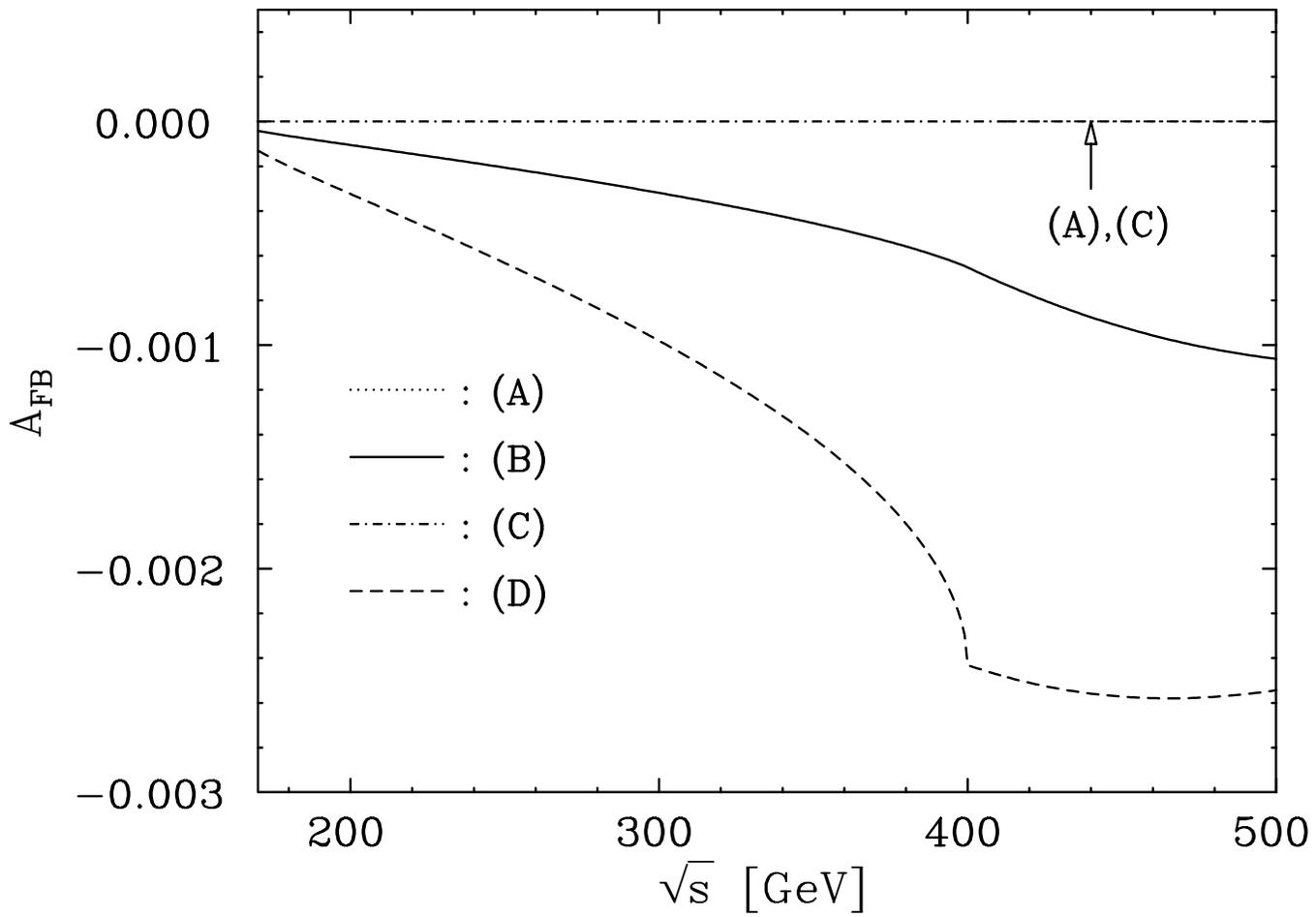
\item[fig.4]
 Forward-backward asymmetry
 \[ A_{FB} =
   \frac{\int ^{0.98}_{0}d\cos \Theta \frac{d\sigma }{d\cos \Theta }
             -\int ^{0}_{-0.98}d\cos \Theta \frac{d\sigma }{d\cos \Theta }}
        {\int ^{0.98}_{0}d\cos \Theta \frac{d\sigma }{d\cos \Theta }
             +\int ^{0}_{-0.98}d\cos \Theta \frac{d\sigma }{d\cos \Theta }}
 \]
 for
 ${\rm e}^{+}_{\rm L}{\rm e}^{-}_{\rm R}\! \rightarrow {\rm W}^{+}{\rm W}^{-}$
 as a function of $\sqrt{s}$.\\
 The values of the masses in (A), (B), (C) and (D) are the same as those
 in the correponding curves in fig.2.
\end{description}

\newpage
\begin{thebibliography}{99}
\bibitem{KL}
D.~C.~Kennedy and B.~W.~Lynn, {\it Nucl. Phys.} {\bf B322} (1989) 1
\bibitem{PT}
M.~E.~Peskin and T.~Takeuchi, {\it Phys. Rev. Lett.} {\bf 65} (1990) 964;
{\it Phys. Rev.} {\bf D46} (1992) 381
\bibitem{AB}
G.~Altarelli and R.~Barbieri, {\it Phys. Lett.} {\bf B253} (1990) 161; \\
G.~Altarelli, R.~Barbieri and S.~Jadach {\it Nucl. Phys.} {\bf B369} (1992) 3
\bibitem{APLS}
C.~Ahn, M.~E.~Peskin, W.~Lynn and S.~Selipsky, {\it Nucl. Phys.} {\bf B309}
(1988) 221
\bibitem{See-Saw}
M.~Gell-Mann, P.~Ramond and R.~Slansky, in : {\it Supergravity}, \\
eds. P.~V.~Nieuwenhuizen and D.~Z.~Freedman (North Holland, Amsterdam, 1979)
p.315 \\
T.~Yanagida, Horizontal Gauge Symmetry and Masses of Neutrinos, \\
in: {\it Proc. Workshop on Unified theories and baryon number in the universe}
(KEK, February 1979), eds. O.~Sawada and A.~Sugamoto (KEK-79-18) p.95 \\
R.~Mohapatra and G.~Senjanovic, {\it Phys. Rev. Lett.} {\bf 44} (1980) 912;
{\it Phys. Rev.} {\bf D23} (1981) 165
\bibitem{HP}
C.~T.~Hill and E.~A.~Paschos, {\it Phys. Lett.} {\bf B241} (1990) 96
\bibitem{BS}
S.~Bertolini and A.~Sirlin, {\it Phys. Lett.} {\bf B257} (1991) 179
\bibitem{GT}
E.~Gates and J.~Terning, {\it Phys. Rev. Lett.} {\bf 67} (1991) 1840
\bibitem{Neg-ST}
D.~C.~Kennedy and P.~Langacker, {\it Phys. Rev. Lett.} {\bf 65} (1990) 2967;
{\bf 66} (1991) 395(E); \\
W.~J.~Marciano and J.~Rosner, {\it Phys. Rev. Lett.} {\bf 65} (1990) 2963
\bibitem{HPZH}
K.~Hagiwara, R.~D.~Peccei, D.~Zeppenfeld and K.~Hikasa, {\it Nucl. Phys.}
{\bf B282} (1987) 253
\bibitem{H}
B.~Holdom, {\it Pyhs. Lett.} {\bf B258} (1991) 156
\bibitem{AW}
T.~Appelquist and G.H.~Wu, {\it Phys. Rev.} {\bf D48} (1993) 3235
\bibitem{PV}
G.~Passarino and M.~Veltman, {\it Nucl. Phys.} {\bf B160} (1979) 151
\bibitem{M}
M.~Marui, preprint OCHA-PP 35 (1993)
\bibitem{FSKO}
J. Fujimoto, Y. Shimizu, K. Kato and Y. Oyanagi, {\it Progr. Theor. Phys.}
{\bf 87} (1992) 1233
\end{thebibliography}
\end{document}